\documentclass[runningheads, a4paper,11pt]{llncs}
\usepackage{etex}
\usepackage{makeidx} % allows for indexgeneration

\usepackage{graphicx}

\usepackage{rotating}
\usepackage{tikz}
\usetikzlibrary{shapes,arrows}
\usetikzlibrary{calc}
\usepackage{msc} 
\usepackage{afterpage}
\usepackage{braket}
\usepackage{amssymb,amsmath}
\usepackage{mathabx}
\usepackage{algorithm}
\usepackage{appendix}
\usepackage{subfig}
\usepackage{psfrag}
\usepackage{amsbsy}
\usepackage{mathrsfs}
\usepackage{dsfont}
\usepackage{url}
\usepackage{listings}
\usepackage{textcomp}
\usepackage{microtype}
\usepackage{paralist} 
\usepackage{booktabs}
\usepackage{extarrows}
\usepackage{comment}
\usepackage{array,arydshln}
\usepackage{siunitx}
\usepackage[utf8]{inputenc}
\usepackage{amsmath,mathtools,amssymb}
\usepackage[framemethod=tikz]{mdframed}
\usepackage{enumitem}
\usepackage{subfig}
\usepackage[letterpaper,hmargin=1.6in,vmargin=1.6in]{geometry}

\normalsize

\usepackage[colorinlistoftodos]{todonotes}

\tikzstyle{connector} = [->,thick]
\tikzstyle{snakeline} = [connector, decorate, decoration={pre length=0.2cm,
                         post length=0.2cm, snake, amplitude=.4mm,
                         segment length=2mm},thick, magenta, ->]

\newcommand*\circled[1]{\tikz[baseline=(char.base)]{\node[shape=circle,draw,minimum size=.05cm] (char) {#1};}}

\newcounter{protocol}
\newenvironment{protocol}[3]{
    \refstepcounter{protocol}%
    \begin{mdframed}[hidealllines=true,backgroundcolor= gray!20]
        \parindent=0pt
        \parskip=2pt
        \begin{center}
            \textbf{Protocol \theprotocol}. #1
        \end{center}
        \medskip
        \textsf{INPUT:} #2
        
        \textsf{OUTPUT:} #3
        \medskip
        }{
    \end{mdframed}    
}

\newcommand{\keywords}[1]{\par\addvspace\baselineskip
\noindent\keywordname\enspace\ignorespaces#1}

{\bfseries}{\itshape}

\newcommand{\aysajan}[1]{\textcolor{blue}{\textbf{NOTE Aysajan: \emph{#1}}}}

\title{Entanglement-based Mutual Quantum Distance Bounding}

\author{}
\institute{}
\author{Aysajan Abidin$^1$, Karim Eldefrawy$^2$, Dave Singel\'ee$^1$}

\institute{$^1$imec-COSIC KU Leuven, Belgium\\
$^2$SRI International, USA}
\begin{document}
\maketitle

\begin{abstract}
Mutual distance bounding (DB) protocols enable two distrusting parties to establish an upper-bound on the distance between them. DB has been so far mainly considered in classical settings and for classical applications, especially in wireless settings, e.g., to prevent relay attacks in wireless authentication and access control systems, and for secure localization. While recent research has started exploring DB in quantum settings, all current quantum DB (QDB) protocols employ quantum-bits (qubits) in the rapid-bit exchange phase, and only perform one-way DB. Specifically, the latest QDB proposals improve initial ones by adding resistance to photon number splitting attacks, and improving round complexity by avoiding communication from the prover to  verifier in the last authentication phase.

This paper presents two new QDB protocols that differ from previously proposed protocols in several aspects: (1) to the best of our knowledge, our protocols are the first to utilize entangled qubits in the rapid-bit exchange phase, previous protocols relied on sending individual qubits, not those from a pair of entangled ones; (2) our second protocol can perform mutual QDB between two parties in one execution, previous QDB protocols had to be executed twice with the prover and verifier roles reversed in each execution; (3) the use of entangled qubits in our protocols thwarts attacks that previous QDB protocols were prone to; (4) and finally, our protocols also eliminate the need for communication from the prover to the verifier in the last authentication phase, which was necessary in some previous QDB protocols. Our work paves the way for several interesting research directions which we briefly discuss in detail in the appendix.
%, e.g., generalizing QDB to group settings, which remains currently open.
%, unlike the classical case.
\end{abstract}

\section{Introduction}\label{Introduction}

%%%%%%%%%%%%%%%%%%%%%%%%%%%%%%%%%%%%%%%%%%%%%%%%%%%%%%
%%%%%%%%%%%%%%%%%%%%%%%%%%%%%%%%%%%%%%%%%%%%%%%%%%%%%%
%
%
%   TODO: 
%   * (Aysajan) Put in LNCS format
%   * (Karim) Motivate QDB and provide an application

%   * (Aysajan) Quantitative comparison table 

%   * (Karim) Explain the benefit of using entanglement:
    
%   * device independent security
    
%   * strong correlation if attacker just reflects

%   * (Dave) Explain why the comment on generating two particle encoding the same value, say, 0, does not make sense -> I now understand where this comment comes from. We argue that the use of entangled qubits offers better security, as it protects the verifier from chosing the challenge value in advance. However, the prover has no means to verify that the challenge it received, is actually an entangled particle and not an un-entangled particle. Therefore, nothing prevents a verifier that wants to cheat, from just generating two un-entangled particles enconding the same value. Or in other words, our argument that entanglement offers stronger security, does not fully hold. It only prevents reflection attacks where the prover reflects the entangled particle instead of sending a normal qubit as response.

%   * (Dave) Update distance fraud attack for mutual DB    
%   * (Dave) attacker (Party A) can always cheat whenever register b and c are equal  
%   * (round_success_probability)^HD(b,c)
%
%
%%%%%%%%%%%%%%%%%%%%%%%%%%%%%%%%%%%%%%%%%%%%%%%%%%%%%%
%%%%%%%%%%%%%%%%%%%%%%%%%%%%%%%%%%%%%%%%%%%%%%%%%%%%%%

Distance Bounding (DB) protocols are cryptographic protocols that combine entity authentication and proximity verification. These protocols enable a verifier to establish an upper-bound on the distance to an untrusted prover. DB was introduced by Brands-Chaum~\cite{BC93} as a primitive to prevent relay attacks on Automatic Teller Machines (ATM) systems. Following this initial proposal of Brands-Chaum, several new DB protocols were proposed, and also implemented~\cite{Rasmussen:2010:RRD:1929820.1929854} and experimentally evaluated. More background on classical DB protocols and their main design blueprints can be found in appendix.

All DB protocols proposed in the literature so far, require an (unpredictable) rapid exchange of bits, i.e., a sequence of fast challenge-response phases. The security of this primitive, and hence also the DB protocol itself, relies on the laws of physics, i.e., that the speed of light is an upper bound on the speed of electromagnetic waves. In RF-based DB protocols, this means in practice that adversaries cannot transmit signals faster than the speed of light, and cannot force signals to arrive faster at the receiver than the actual propagation time to travel the distance to the receiver. This physical law is used to establish an upper bound on the (physical) distance between the prover and the verifier. It is worth noting that recent research \cite{Gerault'19} has questioned certain assumptions (and ``folklore'' design guidelines) and adversarial models in the DB research literature, e.g., that only single bits should be transmitted in the fast-bit exchange phase; we do not tackle such advanced issues in this work. 

%In general, there are two main blueprints of (classical) DB protocols, the so-called Brands-Chaum~\cite{BC93} blueprint or the Hancke-Kuhn~\cite{HK05} one. Both blueprints require an (unpredictable) rapid-bit exchange typically executed after a cryptographic initialization   phase to distribute some randomness. This initial randomness is then used in the sequence of fast challenge-response phases  where security thereof relies on the laws of physics, i.e., that the speed of light is an upper bound on the speed of electromagnetic waves. For example, in RF-based DB protocols adversaries cannot transmit signals faster than the speed of light, this physical fact is used to establish an upper bound on the (physical) distance between the prover and the verifier. More background on classical DB protocols can be found in appendix. 

Up to a decade ago, all DB protocols focused on RF-based DB. A couple of recent proposals \cite{Abidin2019,AbidinQDB} explored the possibility of developing Quantum DB (QDB) protocols. Such QDB protocols rely on principles of quantum physics, namely, that unlike classical bits sent over classical communication channels, quantum bits (qubits) cannot be measured without modifying their states. In particular, if a qubit is in an unknown state, an adversary cannot extract any information about it without destroying it. We review such QDB protocols in Section \ref{Background}. While such protocols already realize some practical versions of one-way QDB, there are still a lot of basic open questions and work to be performed in this emerging research topic; there are also several unexplored areas in the possible protocol design landscape. \\

\textbf{Applications of QDB.} There are a couple of emerging (quantum communication related) applications that could benefit from QDB. The first application is enhancing quantum key distribution (QKD) by augmenting it with QDB. In recent years, QKD systems have been experimentally
evaluated over terrestrial \cite{Honjo:08,Takesue:10,Ribordy_2000} and satellite systems \cite{Liao2017}; to further extend distances over which QKD can operate, such systems could be augmented by quantum relays or repeaters \cite{collins2003quantum,qrepeater} which seem  likely to be constructed in the coming years. We argue that QDB could be added to key establishment functionalities in such systems to obtain distance-bounded mutually shared keys where both the security of the derived keys, and bounds on the distance of  entities such keys are established with, are based on quantum/physical properties. 
Another closely-related application that we envision is to augment systems used to distribute entanglement in what has been termed the ``quantum Internet"~\cite{Wehnereaam9288} with distance bounding guarantees. We acknowledge that this topic (quantum Internet) is still at a very early stage and may be even regarded by some as impractical, but we stress that experimental demonstrations for aspects of it justify why it is physically realizable and interesting to explore. What is unclear is whether such entanglement distribution systems will have practical applications in the near future, but this is out of scope of this paper. We just point out that QDB may be useful in explorations of that topic. \\

\textbf{Contributions.} In this paper we make the following contributions:
(1) We present two new QDB protocols that (to the best of our knowledge) are the first QDB protocols to utilize entangled qubits in the rapid-bit exchange phase; previous protocols relied on sending individual qubits, not those from a pair of entangled ones. (2) Our second protocol can perform mutual QDB between two parties (acting both as prover and verifier) in one execution, previous QDB protocols had to be executed twice with those roles reversed in each execution. (3) We analyze the security and practicality of our QDB protocols and conclude that they are within reach of  commercially available QKD equipment, and other equipment required for similar experimental setups; both types of equipment have been  tested and verified several times as reported in existing literature.\\

\textbf{Why Use Entanglement in QDB?} This is a natural question given that there are QDB protocols \cite{Abidin2019,AbidinQDB} that utilize un-entangled quantum particles in the rapid-bit exchange phase. The benefits of using entangled particles are:

\vspace{-7pt}
\begin{enumerate}

\item Using entangled particles in QDB enables us to develop a mutual QDB protocol that requires 25\% less rounds (and thus communication) than performing two independent executions of a one-way QDB with the parties assuming reversed roles; there are currently no other proposals for performing mutual QDB protocols.

\item In existing QDB protocols that rely on un-entangled quantum particles, the (pseudo) random challenge is first classically produced by the verifier and then encoded as a quantum particle. When entangled particles are used in our QDB protocols, such a (physically) random  challenge remains unknown to an honest verifier (and thus also the adversary that may compromise it during protocol execution) until later when measured; this provides less room for an adversary to cheat in the challenge generation step. We view this property as an enabler for \emph{device independent security} because the processing device used in QDB cannot bias the randomness used in the protocol in this case.

\item Our QDB protocols can detect a new class of attacks, where a cheating prover immediately reflects challenge bits back without processing them in an attempt to shorten the distance perceived by the verifier. When entangled particles are used, (unusually) strong correlations between the particles retained by the verifier and those reflected by prover -- pretending to be legitimate responses to the verifier's challenges -- can be detected at the verifier.

\end{enumerate}

\noindent Finally, we note that our protocols would still work if the entangled quantum particles (qubits) are replaced with un-entangled ones, but in this case the security of our protocols reduces to the security of the pseudorandom number generator (PRNG) used to generate the (pseudo) random challenge bits.\\  

%\dave{Our security argument does not fully hold, it only applies in theory. In practice, the prover has no way to verify that a (cheating) verifier actually has sent an entangled particle.}

\textbf{Paper Outline.} The rest of this paper is organized as 
follows, Section \ref{Background} reviews preliminaries required for this paper and briefly discusses related work. Section \ref{Proposal} contains the main novel material in this paper, it introduces two new entanglement-based QDB protocols, one that can only perform one-way DB, while the other performs mutual DB. Section \ref{SecurityAnalysis} contains the security analysis of the proposed QDB protocols. Section \ref{Future} presents open issues and future work, and Section \ref{Conclusion} concludes the paper.

\section{Background and Preliminaries} \label{Background}
This section covers the necessary background required for the rest of the paper. We start with a description of classical DB protocols, followed by a brief overview of qubits and quantum entanglement, and then discuss related work in QDB.

%\section{Hancke-Kuhn Distance Bounding Protocol}
%\vspace{-50pt}

\subsection{Classical distance bounding protocols} 
\label{classicalDB-Apndx}

\begin{figure}[htp]
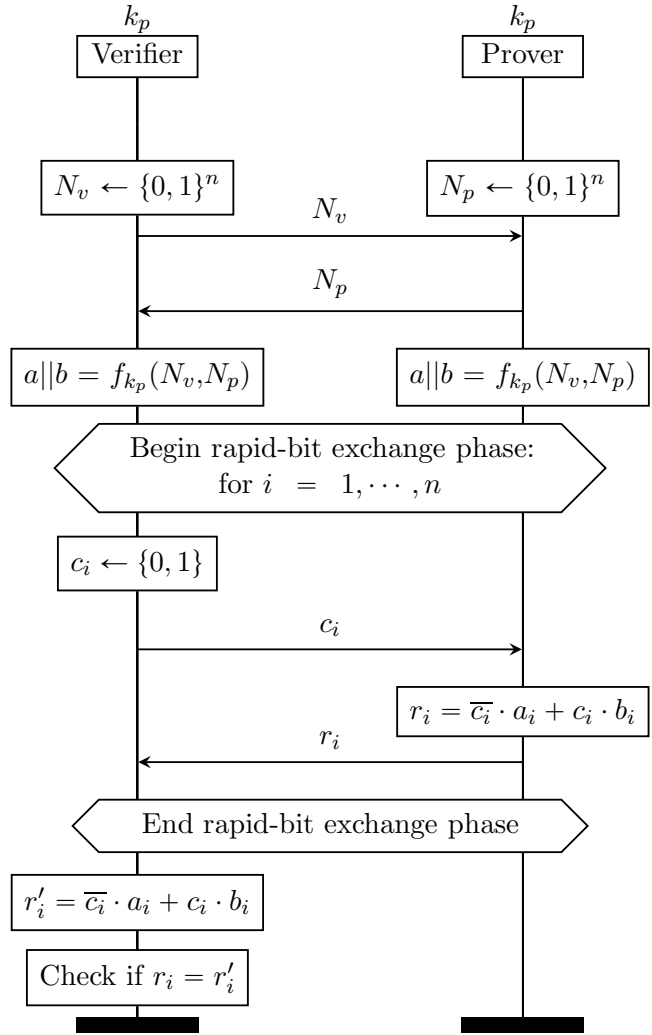

\begin{center}
\setlength{\instdist}{35mm}
\scriptsize
\scalebox{1}{
\begin{msc}{}
\drawframe{no}
\declinst{v}{$k_{p}$}{Verifier}
\declinst{p}{$k_{p}$}{Prover}
\nextlevel[-0.5]
\action*{$N_{v}$ $\leftarrow$ $\{0,1\}^n$}{v}
\action*{$N_{p}$ $\leftarrow$ $\{0,1\}^n$}{p}
\nextlevel[2]
\mess{$N_{v}$}{v}{p}
\nextlevel[2]
\mess{$N_{p}$}{p}{v}
\nextlevel
\action*{$a||b$ = $f_{k_p}$($N_{v}$,$N_{p}$)}{v}
\action*{$a||b$ = $f_{k_p}$($N_{v}$,$N_{p}$)}{p}
\nextlevel[2]
\condition{Begin rapid-bit exchange phase:\\ for $i=1,\cdots,n$}{v,p}
\nextlevel[3]
\action*{$c_i\leftarrow\{0,1\}$}{v}
\nextlevel[3]
\mess{$c_i$}{v}{p}
\nextlevel[1]
\action*{$r_i = \overline{c_i} \cdot a_i + {c_i} \cdot {b_i}$}{p}
\nextlevel[2]
\mess{$r_i$}{p}{v}
\nextlevel[1]
\condition{End rapid-bit exchange phase}{v,p}
\nextlevel[2]
\action*{$r'_i = \overline{c_i} \cdot a_i + {c_i} \cdot {b_i}$}{v}
\nextlevel[2]
\action*{Check if $r_i = r'_i$}{v}
\nextlevel
\end{msc}
}
\caption{Hancke-Kuhn Distance Bounding Protocol \cite{HK05}}
\label{fig:basic-db}
\end{center}
\end{figure}

DB protocols \cite{BC93} allow one 
entity (verifier) to obtain an upper-bound on the distance to 
another entity (prover), in addition to authenticating the latter.
Figure \ref{fig:basic-db} shows an example of a generic DB protocol: the Hancke-Kuhn protocol \cite{HK05}, where $k_p$ is a shared key between the prover and the verifier, $f$ is a pseudo-random function, and $a$ and $b$ are of length $n$-bit each. The core of any \textit{one-way DB} protocol is the distance measurement phase, whereby the verifier measures round-trip time between sending its challenge and receiving the reply from the prover. The verifier's challenges are unpredictable to the prover and replies are computed as a function of these challenges. Thus, the prover cannot reply to the verifier sooner than it received the challenges. The prover, therefore, cannot pretend to be  closer to the verifier than it really is (only further). 

%The verifier and the prover each generate $n$ $b$-bit nonces $c_i$ and $r_i$ ($0<~i\le~n$), respectively. In the first (Brands-Chaum) DB protocol \cite{BC93}, the prover also commits to its nonces. The verifier sends all $c_i$ to the prover, one at a time. Once each $c_i$ is received, the prover computes, and responds with a function computed on its own nonce and that of the verifier, $f(c_i,r_i)$. The verifier checks the reply and measures the elapsed time between each challenge and response. The process is repeated $n$ times and the protocol completes successfully only if \textit{all} $n$ rounds succeed and all responses correspond to the prover's committed value. The processing time on the prover's side $\alpha=t_{s}^P-t_{r}^P$ must be negligible (compared to time of flight of the signal); otherwise, a computationally powerful prover could claim a false bound. This time might be tolerably small, depending on the underlying technology, the distance measured and required security guarantees. 

The first (Brands-Chaum) DB protocol \cite{BC93} comprises three phases, namely, an initialisation phase, a rapid-bit exchange phase consisting of $n$ rounds, and an authentication phase. In the initialisation phase, the prover first commits to a randomly generated $n$-bit nonce $N$. Then in the $i$-th round of the rapid-bit exchange phase, for $i=1,\cdots,n$, the verifier sends a random challenge bit $c_i$ to the prover, and the prover computes and responds with $r_i=N_i\oplus c_i$. In the last phase, the verifier checks the reply and measures the elapsed time between each challenge and response. The protocol completes successfully only if \textit{all} $n$ rounds succeed and all responses correspond to the prover's committed value (i.e., $c_i\oplus r_i=N_i, \,i=1,\cdots,n$). The processing time on the prover's side $\alpha=t_{s}^P-t_{r}^P$ must be negligible (compared to time of flight of the signal); otherwise, a computationally powerful prover could claim a false bound. This time might be tolerably small, depending on the underlying technology, the distance measured and required security guarantees. 

The security of DB protocols relies on two assumptions: (1) challenges are random and unpredictable to the prover before being sent by the verifier; (2) challenges traverse the distance between the prover and the verifier at maximum possible speed, i.e., the speed of electromagnetic waves. After executing the DB protocol, the verifier knows that the distance to the prover is at most $\frac{t_r^V - t_s^V-\alpha}{2}\cdot c$, where $\alpha$ is the processing time of the prover (ideally,  negligible) and $c$ is the speed of 
light \cite{BC93}. DB protocols typically require $(2\cdot n + {\cal C})$ messages, where ${\cal C}$ is the number of 
messages exchanged in the pre- and post-processing protocol phases. 
Typically, ${\cal C} << n$ and thus can be ignored.

In some cases (e.g., distributed localization), there is a need for mutual DB between $P_1$ and $P_2$. This can be achieved by modifying the one-way DB protocol such that each response from $P_2$ to a challenge by $P_1$ also includes a challenge from $P_2$ to $P_1$. This requires $2n + 2{\cal C} + 1$ messages instead of $2(2\cdot n + {\cal C})$ for mutual DB and is shown in \cite{CBH03}. Both parties generate and commit to two random bit 
strings $[c_1,c_2,...,c_n]$ and $[s_1,s_2,...,s_n]$. $P_1$ starts by sending the first challenge 
bit $c_1$ and $P_2$ replies with 
$c_1 \oplus s_1$. $P_1$ measures the time between sending $c_1$ and receiving the 
response. $P_1$ then replies with $c_2 \oplus s_1$. $P_2$ measures
the time between sending $c_1 \oplus s_1$ and receiving the response. This process is repeated 
$n$ times. The mutual DB procedure is considered successful if both parties verify all 
responses and match previously committed values (see \cite{CBH03} for more details).

\begin{table}[h]
\centering
\caption{A rule for encoding classical bits as qubits.}
\resizebox{1\textwidth}{!}{\begin{minipage}{\textwidth}
\begin{tabular}{|c|c|c|}
\hline
\textbf{Data} & \textbf{Comp. (or $+$) basis} & \textbf{Hadamard (or $\times$) basis} \\
\hline
~0~ & $\Ket{0}$ (i.e., $\rightarrow$) & $\Ket{+}$ (i.e., $\nearrow$)\\
\hline
~1~ & $\Ket{1}$ (i.e., $\uparrow$) & $\Ket{-}$ (i.e., $\nwarrow$)\\
\hline
\end{tabular}\label{tab1}
\end{minipage}}
\end{table} 
%\vspace{-1cm}

\subsection{Qubits}
\vspace{-5pt}
\begin{table}[htp]
\centering
\caption{A rule for encoding classical bits as qubits.}
\resizebox{1\textwidth}{!}{\begin{minipage}{\textwidth}
\centering
\begin{tabular}{|c|c|c|}
\hline
\textbf{Data} & \textbf{Comp. (or $+$) basis} & \textbf{Hadamard (or $\times$) basis} \\
\hline
~0~ & $\Ket{0}$ (i.e., $\rightarrow$) & $\Ket{+}$ (i.e., $\nearrow$)\\
\hline
~1~ & $\Ket{1}$ (i.e., $\uparrow$) & $\Ket{-}$ (i.e., $\nwarrow$)\\
\hline
\end{tabular}\label{tab1}
\end{minipage}}
\end{table} 
%\vspace{-1cm}

%\vspace{-12pt}

A qubit is a unit of quantum information, just as a bit (0 or 1) is the classical unit of information. A qubit is a vector in a 2-dimensional Hilbert space (a vector space with inner product). The basis $$\{\Ket{0}=\begin{bmatrix} 1 \\ 0
\end{bmatrix},~\Ket{1}=\begin{bmatrix} 0 \\ 1
\end{bmatrix}\}$$ for a qubit is called the computational basis, whereas the basis $$\{\Ket{+}=(\Ket{0}+\Ket{1})/\sqrt{2},~\Ket{-}=(\Ket{0}-\Ket{1})/\sqrt{2}\}$$ is called the diagonal (or the Hadamard) basis. In general, a normalised quantum state can be expressed as a superposition of $\Ket{0}$ and $\Ket{1}$ as 
$$\alpha\Ket{0}+\beta\Ket{1},$$ where $\alpha,\beta\in\mathbb{C}$ satisfying $|\alpha|^2+|\beta|^2=1$.

If a qubit in state $\alpha\Ket{0}+\beta\Ket{1}$ is measured in the computational basis, then $\Ket{0}$ is obtained with probability $|\alpha|^2$ and $\Ket{1}$ with probability $|\beta|^2$. And upon measurement in the computational basis, the qubit with original state $\alpha\Ket{0}+\beta\Ket{1}$ collapses into $\Ket{0}$ or $\Ket{1}$, which is different from the original. 

The four states, $\Ket{0}$, $\Ket{1}$, $\Ket{+}$, and $\Ket{-}$, have some nice properties. For example, they satisfy that 
$$\Ket{0}=(\Ket{+}+\Ket{-})/\sqrt{2}$$
and 
$$\Ket{1}=(\Ket{+}-\Ket{-})/\sqrt{2}.$$
If the qubits $\Ket{0}$ and $\Ket{1}$ are measured in the computational basis, then the states are not changed; whereas a measurement in the Hadamard basis completely destroys the state. In the latter case, either $\Ket{+}$ or $\Ket{-}$ is obtained with equal probability. Similarly, if the qubits $\Ket{+}$ and $\Ket{-}$ are measured in the Hadamard basis, then the states do not change; whereas a measurement in the computational basis destroys the state, and either $\Ket{0}$ or $\Ket{1}$ is obtained with equal probability. It is this principle that is used in \cite{Abidin2019,AbidinQDB}, while in the present paper we use it together with quantum entanglement (cf. Section \ref{sec:entanglement}).

These four states correspond to different polarisations of photons. The states $\Ket{0}$ and $\Ket{1}$ correspond to horizontally $\rightarrow$ and vertically $\uparrow$ polarised photons, respectively, whereas the states $\Ket{+}$ and $\Ket{-}$ to $\nearrow$ (45$^\circ$) and $\nwarrow$ (-45$^\circ$) polarised photons. 
The classical bit value 0 is encoded as a qubit in state $\Ket{0}$ or $\Ket{+}$, and the value 1 is encoded as a qubit in state $\Ket{1}$ or $\Ket{-}$. The qubits are measured either in the computational or the Hadamard basis. Throughout this paper, ($+$) denotes the computational basis and ($\times$) the Hadamard ($\times$) basis; see Table~\ref{tab1}. 

If 0 is encoded as $\rightarrow$ polarized photon, it can be decoded correctly as 0 only in the $+$ basis, whereas if 0 is encoded as $\nearrow$ polarized photon, then it can be decoded correctly as 0 only in the $\times$ basis. This also applies to the case when a 1 is encoded in the $+$ basis and the $\times$ basis, respectively. In any of these cases, if a polarized photon is decoded using a wrong basis, then one obtains a random bit. Therefore, by using the above encoding method one can send information encoded as polarized photons so that no one can copy or read reliably without knowing the bases used for encoding.

Later in the paper, we use $\Ket{x}_y$, for $x,y\in\{0,1\}$, to denote encoding of a classical bit $x$ into a qubit using the basis determined by $y$. 

\subsection{Quantum Entanglement}
\label{sec:entanglement}
What we have learnt in the previous subsection is that if we measure a qubit in an unknown state, not only do we obtain a purely  random result, but we destroy the original state of the qubit. Another useful property of qubits is that they can be \emph{entangled}.  A two-qubit state $\Ket{\beta}$ is called \emph{entangled} if it cannot be expressed as a (tensor) product of its component states. For example, the following four two-qubit states (a.k.a., Bell states or Einstein-Podolsky-Rosen (EPR) pairs)
$$
\begin{aligned}
\Ket{\beta_{00}}&=\frac1{\sqrt{2}}(\Ket{00}+\Ket{11}),\\
\Ket{\beta_{01}}&=\frac1{\sqrt{2}}(\Ket{00}-\Ket{11}),\\
\Ket{\beta_{10}}&=\frac1{\sqrt{2}}(\Ket{10}+\Ket{01}), \mbox{ and}\\
\Ket{\beta_{11}}&=\frac1{\sqrt{2}}(\Ket{01}-\Ket{10})
\end{aligned}
$$
are (maximally) entangled states that are mutually orthogonal. 
The fact that entanglement sources generate pairs of entangled qubits is relevant and could be explained more prominently.
Entangled qubits exhibit strong correlations with no classical analog.  We use entanglement in our protocols as follows. When an entanglement source (e.g., one emitting pairs of spin $\frac1{2}$ particles) generates a pair of entangled qubits (say, on the verfier side), one half is sent to the prover and the other half is kept locally. Then both prover and verifier measure their half of the entangled qubits in the same measurement setting. So after learning its local measurement outcome, the verifier can predict with probability 1 what would the prover obtain if it also measures its half of the entangled qubits in the same setting. The prover will then send as a response its measurement outcome encoded into a state of a new qubit (cf. Section \ref{sec:new-one-way-QDB}). Prior to their measurements, neither of the prover and verifier can predict the value of the challenge or response. In other words, there is no a priori information encoded into the sate of entangled qubits. The information comes into existence only after measurement. 

\subsection{Related Work in QDB} \label{Rel-Work}
\begin{figure}[htp]
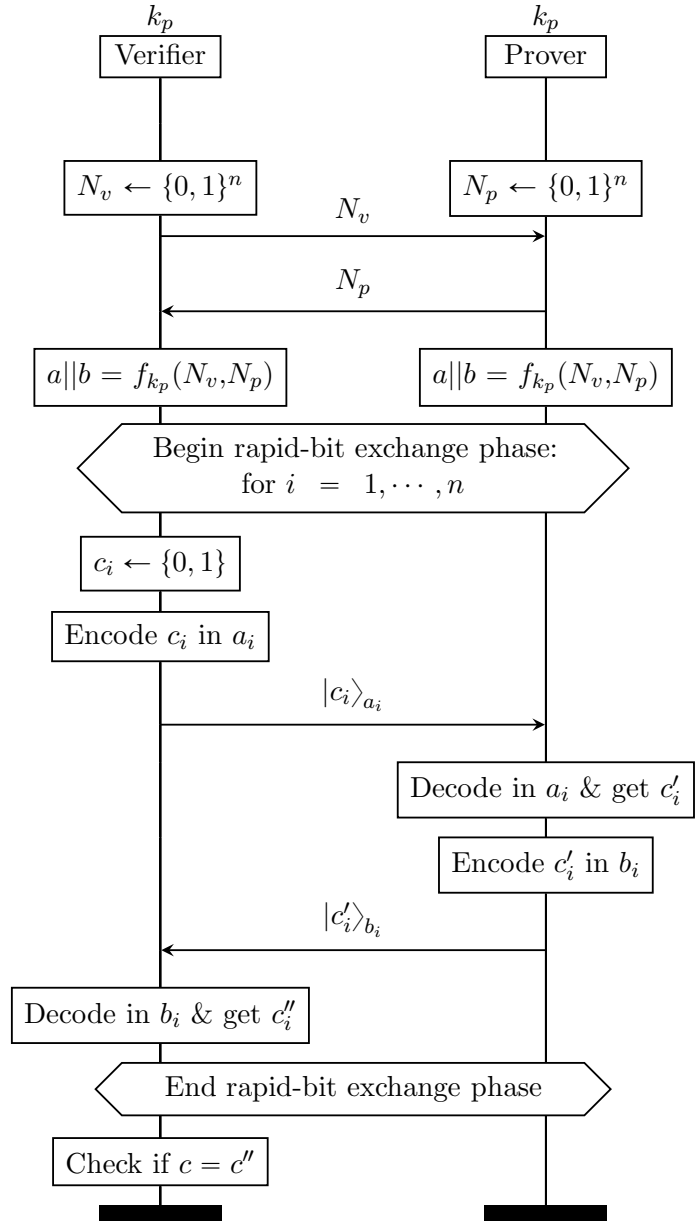

\begin{center}
\setlength{\instdist}{35mm}
\scriptsize
\scalebox{1}{
\begin{msc}{}
\drawframe{no}
\declinst{v}{$k_{p}$}{Verifier}
\declinst{p}{$k_{p}$}{Prover}
\nextlevel[-0.5]
\action*{$N_{v}$ $\leftarrow$ $\{0,1\}^n$}{v}
\action*{$N_{p}$ $\leftarrow$ $\{0,1\}^n$}{p}
\nextlevel[2]
\mess{$N_{v}$}{v}{p}
\nextlevel[2]
\mess{$N_{p}$}{p}{v}
\nextlevel
\action*{$a||b$ = $f_{k_p}$($N_{v}$,$N_{p}$)}{v}
\action*{$a||b$ = $f_{k_p}$($N_{v}$,$N_{p}$)}{p}
\nextlevel[2]
\condition{Begin rapid-bit exchange phase:\\ for $i=1,\cdots,n$}{v,p}
\nextlevel[3]
\action*{$c_i\leftarrow\{0,1\}$}{v}
\nextlevel[2]
\action*{Encode $c_i$ in $a_i$}{v}
\nextlevel[3]
\mess{$\Ket{c_i}_{a_i}$}{v}{p}
\nextlevel[1]
\action*{Decode in $a_i$ \& get $c'_i$}{p}
\nextlevel[2]
\action*{Encode $c'_i$ in $b_i$}{p}
\nextlevel[3]
\mess{$\Ket{c'_i}_{b_i}$}{p}{v}
\nextlevel[1]
\action*{Decode in $b_{i}$ \& get $c''_i$}{v}
\nextlevel[2]
\condition{End rapid-bit exchange phase}{v,p}
\nextlevel[2]
\action*{Check if $c=c''$}{v}
\nextlevel
\end{msc}
}
\caption{The (one-way) QDB protocol from \cite{Abidin2019}.}
\label{fig:QDB}
\end{center}
\end{figure}

The most relevant QDB protocol is the one proposed in \cite{Abidin2019} and illustrated in Figure \ref{fig:QDB}. In that protocol  $k_p$ is a shared secret key between the prover and the verifier. Both parties generated random values $N_v$ and $N_p$ and exchange them, and a keyed pseudo-random function (PRF) $f_{k_p}$ is applied to $N_v$ and $N_p$  to generate a random bit sequence that is parsed as two equal sized registers $a$ and $b$. We denote by $x_i$ the bit at position $i$ of register $x$ where $x \in \{a,b\}$.
In the $i$-th round of rapid-bit exchange, the verifier generates a random challenge bit $c_i$ and encodes it in a basis determined by $a_{i}$. The prover decodes the challenge qubit in a basis determined by $a_i$ and obtains $c'$ and re-encodes it (as a response qubit) in a basis determined by $b_{i}$. If, in the $i$-th round of the rapid-bit exchange phase, the verifier obtains the same challenge bit by decoding the prover's response qubit in the basis determined by bit $b_{i}$, then  verifier can deduce that it is the prover which sent this response. This is because the only party that can correctly decode the challenge qubit without errors is the prover, i.e., only the correct prover can re-encode the challenge bit in $b_{i}$. Once the rapid-bit exchange is completed, the verifier checks whether the decoded bits $c^{''}_i$ match the generated challenge bits $c_i$.\\

\begin{comment}
\begin{protocol}{\texttt{(One-way) QDB protocol from \cite{Abidin2019}  (see Figure \ref{fig:QDB}).} \label{prot:ow-qdb-rel-work}}%
    {a shared key $k_p$}
    {a bit $b$ indicating whether $c=c''$}
\begin{enumerate}
    \item The verifier (V) and prover (P) generate two random $n$-bit strings $N_{v}$ and $N_{p}$
    \item V and P exchange $N_{v}$ and $N_{p}$
	\item V and P generate a $2n$ long random string $a||b$ = $f_{k_p}$($N_{v}$,$N_{p}$, and split it into two equal substrings $a$ and $b$.
	\item Rapid-bit Exchange Phase: for $i=1,\cdots,n$
	\begin{enumerate}
	    \item V generates a random bit $c_i$ and encodes it as a qubit ($\Ket{c_i}_{a_i}$) in a basis determined by the bit $a_i$
	    \item V sends the encoded qubit $\Ket{c_i}_{a_i}$ to P
	    \item P receives $\Ket{c_i}_{a_i}$ and decodes it in a basis determined by $a_i$. Denote the decoded bit as $c'_i$.
	    \item $c'$ is encoded as a qubit ($\Ket{c'_i}_{b_i}$) in a basis determined by the bit $b_i$.
	    \item P sends $\Ket{c'_i}_{b_i}$to V who decodes it in the basis determined by $b_i$. Denote the decoded bit as $c''_i$.
    \end{enumerate}
    \item V output $b=1$ if $c = c''$ and  $b=0$ otherwise.
\end{enumerate}
\end{protocol}

\end{comment}

The protocol in \cite{Abidin2019} is an improvement to an earlier work in \cite{AbidinQDB} where the authors present the first QDB protocol that falls under the Brands-Chaum blueprint, in that it requires an additional authentication phase in the end. More specifically, the protocol in \cite{AbidinQDB} works as follows. Again, let $k_p$ be a shared key between the prover and verifier. In the initialisation phase, both parties exchange randomly generated values $N_v$ and $N_p$, and compute $a=f_{k_p}(N_v,N_p)$ of, say, length $2n$. Then, in the $i$-th rapid-bit exchange phase, for $i=1,2,\cdots,n$, the verifier encodes a randomly generated challenge bit $c_i$ in a basis determined by $a_{2i-1}$ and sends the resulting qubit to the prover. The prover decodes the received challenge qubit in the basis determined by $a_{2i-1}$, obtains $c_i'$, re-encodes it in the basis determined by $a_{2i}$, and send the resulting qubit as its response. Upon receiving the reponse qubit, the verifier decodes it in the basis determined by $a_{2i}$ and obtains $c_i''$. In the last phase of the protocol, the prover sends a MAC (message authentication code) tag for $(ID_p,ID_v,N_p,N_v,c_1',\cdots,c_n')$, where $ID_p$ and $ID_v$ stands for the identity of the prover and verifier, respectively, computed using the shared key $k_p$ to the verifier, which verifies the MAC tag by comparing it to a MAC tag for $(ID_p,ID_v,N_p,N_v,c_1'',\cdots,c_n'')$ that it computes itself. 

Lately, Abidin proposed a hybrid (one-way) DB protocol that uses classical bits for challenges and qubits for responses in \cite{Abidin2020}. 

Although these protocols use qubits in the rapid-bit exchange phase, they do not utilise entangled qubits. In fact, it is not immediately clear whether we can expand the design space for QDB using entangled qubits. One of the main motivations behind the current paper is to investigate whether entangled qubits can be utilized to design a QDB protocol. To this end, we propose two new QDB protocols employing entangled qubits. 
%\textcolor{blue}{TODO: Briefly describe the other/older QDB protocol in \cite{AbidinQDB},  clearly state that none use entangled particles/qubits, which provides some advantages, but is also an interesting unexplored part of the design space of QDB.}

\section{Entanglement-based QDB} \label{Proposal}

In this section we present details of the new entanglement-based QDB protocols.
We start with a brief discussion of the intuition behind our new protocols.

\subsection{Intuition of Entanglement-based QDB}
As is standard in most DB literature, we  assume that the two parties engaging on the protocol have a shared cryptographic key that can be used together with a PRF to generate a random bit string that can be split into several sub-strings which we call registers. In the one-way DB case, we require this shared random binary string to be (deterministically) split into two registers $a$ and $b$; in the mutual QDB case we require it to be split into three registers $a$, $b$, and $c$.
The main idea behind the new QDB protocols is to use entangled particles (denoted by \emph{EP} in the Figures \ref{Protocol1} and \ref{Protocol2}) in the rapid-bit exchange phases.  In the base protocol for one-way QDB, two entangled particles are prepared by a verifier and one of the particles is sent by that verifier to the prover.  Instead of encoding a randomly generated new challenge bit $c$ as a qubit in a basis determined by the random bits in register $a$, as was done in previous work \cite{Abidin2019} (and as illustrated in the one-way protocol in Figure \ref{fig:QDB}), we now generate two entangled particles and determine the basis to measure them, and also the basis for encoding the response, by bits of the two registers $a$ and $b$ in the one-way DB case, and three registers $a$, $b$, and $c$ in the mutual DB case. 

The protocol blueprint that uses two registers only achieves one-way DB though. In theory, one can perform mutual DB by using the one-way DB protocol in Figure \ref{Protocol1}. This is done by executing the protocol twice, with reversed role in the second execution. %If one is only interested in one-way DB, there is no real benefit to encode the response from the prover to the verifier as an entangled particle (i.e. one could also rely on using the QDB protocol in \cite{Abidin2019}). This changes if one is interested in a single protocol performing mutual QDB. In this case, if one changes the response from the prover to verifier to be encoded using entangled particles, then one can essentially use that entangled response also as a challenge in the other direction. As a result, by using entangled particles, one can achieve mutual DB with fewer rounds and communication compared to two sequential (independent) executions of the one-way QDB protocol. 
For a mutual DB with fewer communication rounds, we need the prover's response to the verifier's challenge to be unpredictable by the verifier so that it can also be regarded as a challenge by the verifier. With this in mind, we prepare the prover's response qubit as encoding of the measurement outcome of the entangled qubit XORed with a random bit.  
One subtle issue in this case is that the entanglement-based mutual QDB protocol needs an additional random bit to be committed to (denoted by $r$ in Figure \ref{Protocol2}). This will now be discussed more in detail below in Section \ref{subsec:one-way_QDB_protocol}.

\subsection{Details of Entanglement-based QDB \label{subsec:one-way_QDB_protocol}}
We first present details of our new entanglement-based one-way QDB protocol; we then present the entanglement-based mutual QDB one.

%\textcolor{blue}{TODO: I think that denoting each step in figures  \ref{Protocol1} and \ref{Protocol2} with a step number, and referencing those step numbers in the descriptions in the subsections below will make the protocols clearer to understand. }

\subsubsection{Entanglement-based One-Way QDB \label{sec:new-one-way-QDB}}
An schematic illustration of the  steps of the base one-way QDB protocol is provided in Figure \ref{Protocol1} (top and middle portions). The base one-way QDB protocol enables one party (verifier) to  establish an upper bound on the distance between itself and another party (prover) with a single execution of this protocol. As mentioned before, a single execution of this base one-way DB protocol does not provide the prover with any guarantees on the upper bound on the distance to the verifier. To do this there are two options: (1) the first approach to establish mutual DB is to execute two independent sequential runs of the one-way DB protocol as shown in Figure \ref{Protocol1}. Figure \ref{Protocol1}  shows two such independent runs (middle and bottom sections of the figure) of the protocol with the two parties taking the opposite roles in the second run. In Figure \ref{Protocol1}, Party A acts as a verifier in the first run (middle of the figure) and acts as a prover in the second run (bottom of the figure), and Party B acts as a prover in the first run and acts as a verifier in the second run. (2) The second approach to establish mutual DB is to perform a single execution of the mutual QDB protocol, as will be outlined later in Section \ref{subsec:mutual_QDB_protocol} and Figure \ref{Protocol2}.

In the base one-way QDB protocol, during the initialization phase of the protocol, the two parties exchange two random nonces $N_A$ and $N_B$ and apply the keyed PRF $f_k$ to the exchanged nonces (step 1 and 2 in the figure). The result of the PRF is a random binary string that is then split into two parts which we denote as registers $a$ and $b$. These steps are shown in the top portion of Figure \ref{Protocol1}.  The protocol then proceeds as follows (the sequence of the subsequent steps are repeated $n$ times):
\begin{enumerate}
\setcounter{enumi}{2}
    
    \item In the $i$-th round of rapid-bit exchange, Party A (the verifier) generates a pair of entangled particles EP$_i$, which is one of those mutually orthogonal Bell states $\beta_{00},\,\beta_{01},\,\beta_{10},\,\beta_{11}$ from Section \ref{sec:entanglement} and sends one half, which is again denoted as EP$_i$, to Party B, starts a local timer (clock denoted as CLK in Figure \ref{Protocol1}) when EP$_i$ is sent to the prover.  The verifier then uses $a_i$ to determine the basis in which to measure its half of EP$_i$. Party B (the prover) eventually receives EP$_i$ and measures it in a basis determined by register $a_i$. Denote the measurement result as $m'_i$.
    
    \item Party B encodes the result $m_i'$ of the measurement of the received EP$_i$ in a basis determined by register $b_i$. The encoded result $\Ket{m'_i}$is then sent from Party B to Party A. Next, upon receiving the encoded response, Party A performs the following steps:
    
 \begin{enumerate}   
    \item It stops the clock to measure the time of flight and thus the upper bound on the distance to the prover, and decodes the response in a basis determined by register $b_i$.
    
   \item  It then checks if the response decoded on the basis determined by register $b_i$ is equal to the measurement value of EP$_i$. If this check fails, it aborts the protocol. 
  \end{enumerate}    
  
   \item \textit{Optional:} The protocol can then be repeated with the roles of Parties A and B reversed, as shown in the bottom section (step 5 and step 6) of Figure \ref{Protocol1}.
    
\end{enumerate}

%%%There are a couple of things to note here, first, the prover's response in Figure \ref{Protocol1} does not involve an entangled particle; the  prover's response can be just a quantum particle encoded in a basis determined by register $b$. To simplify the illustration in Figure \ref{Protocol1} we choose not to require prover's response to be an entangled particle. Also, especially that this is is the base one-way DB protocol and such an entangled response is only beneficial when one is interested in mutual DB.

%% DAVE: I would propose to delete the text above (I now put it in a comment, so that is no longer visible in the pdf). The reason why I would delete the text, is because it is very confusing. We present a protocol as an ENTANGLEMENT-based one-way QDB. Then it is a strange message to say that this protocol does not need to use an entangled pair. If we want to write something, then I would say that it is possible to transform the protocol of Figure \ref{Protocol1} into a protocol that does not rely on an entangled pair, but that this is not shown in the paper. 

%\textcolor{blue}{TODO: We need two figures describing the schematic components of a  realization of the protocols in Figures \ref{Protocol1} and \ref{Protocol2}, i.e., schematics similar to Figure 4 in \cite{Abidin2019}.}

\tikzset{meter/.append style={draw, inner sep=10, rectangle, font=\vphantom{A}, minimum width=30, line width=.8,
 path picture={\draw[black] ([shift={(.1,.3)}]path picture bounding box.south west) to[bend left=50] ([shift={(-.1,.3)}]path picture bounding box.south east);\draw[black,-latex] ([shift={(0,.1)}]path picture bounding box.south) -- ([shift={(.3,-.1)}]path picture bounding box.north);}}}

\tikzset{XOR/.style={draw,circle,append after command={
        [shorten >=\pgflinewidth, shorten <=\pgflinewidth,]
        (\tikzlastnode.north) edge (\tikzlastnode.south)
        (\tikzlastnode.east) edge (\tikzlastnode.west)
        }
    }
}

\tikzstyle{decision} = [diamond, draw, fill=blue!20, text width=.5em, text badly centered]

\begin{figure*}[hpt]
\begin{center}
\begin{tikzpicture}[scale=.9,thick,every node/.style={minimum height=.6cm}, squarednode/.style={rectangle, draw, very thick}]
  \scriptsize
  %Initialisation phase
  \node[] at (-4.5,10) (Nv) {$N_A \leftarrow\{0,1\}^\lambda$};
  \node[] at (4,10) (Np) {$N_B \leftarrow\{0,1\}^\lambda$};
  \draw[->] ($(Nv.east)+(1.8,0)$) -- node[above]{{\tiny{\circled{1}}} $N_A$}($(Np.west)+(-2,0)$);
  \draw[<-] ($(Nv.east)+(1.8,-1)$) -- node[above]{{\tiny{\circled{2}}} $N_B$} ($(Np.west)+(-2,-1)$);
  
  \node[] at (-4.5,9) (prf1) {$a||b = f_k(N_A, N_B)$};
  \node[] at (4,9) (prf2) {$a||b = f_k(N_A, N_B)$};
  
  \node[draw] at (3,7.5) (regpa) {\scriptsize{Register $a$}};
  \node[draw] at (5,7.5) (regpb) {\scriptsize{Register $b$}};
  
  \node[draw] at (-5.5,7.5) (regva) {\scriptsize{Register $a$}};
  \node[draw] at (-3.5,7.5) (regvb) {\scriptsize{Register $b$}};
  
  \draw[->] ($(prf1.south) + (-1.3,0)$) to (regva.north);
  \draw[->] ($(prf1.south) + (-.9,0)$) to (regvb.north);
  
  \draw[->] ($(prf2.south) + (-1.3,0)$) to (regpa.north);
  \draw[->] ($(prf2.south) + (-.9,0)$) to (regpb.north);

    %Distance bounding phase
  
  \draw[dotted] (-7,6.75) -- (-2,6.75);
  \draw[dotted] (1.25,6.75) -- (5.75,6.75);
  
    %Define roles: A=V , B=P
  \node[squarednode] at (-6,6) (Nv) {\scriptsize{$A \leftarrow$ Verifier}};
  \node[squarednode] at (5,6) (Np) {\scriptsize{$B \leftarrow$ Prover}};
  
  %Prover to Verifier DB
  \node at (-4.5,5) {For $i=1,2,\cdots,n$};
  
  \node[draw, ellipse, minimum height=1cm, minimum width=2cm] at (-3.5,4) (sourcev) {\scriptsize{Source}};
  \node[meter] (meterv) at (-3.5,2) {};
  \draw[->] (sourcev.south) --node[right]{EP$_i$} (meterv.north);
  %\draw[->] (sourcev.east) -- node[above]{EP} (meterp.west);
  
  \draw (-6,2) node[draw,minimum width=1cm,fill=white](regv1){\scriptsize{Register $a$}};
  \draw[->] (regv1.east) --node[above]{$a_i$} (meterv.west);

  \node[meter] (meterp) at (2.35,3.5) {};
  \draw[->] (sourcev.east) --node[above,sloped]{{\tiny{\circled{3}}} EP$_i$} (meterp.west);
  \draw ($(sourcev.east)+(0.2, -.55)$) node {\begin{tabular}{l}\scriptsize{start}\\\scriptsize{CLK}\end{tabular}};%
  
  \draw (4.85,3.5) node[draw,minimum width=1cm,fill=white](regp1){\scriptsize{Register $a$}};
  \draw[->] (regp1.west) --node[above]{$a_i$} (meterp.east);

  \draw (4.85,-.5) node[draw,minimum width=1cm,fill=white](regp2){\scriptsize{Register $b$}};
  \draw (2.35,-.5)node[draw,minimum width=1cm,fill=white](Pp) {\scriptsize{Encoder}};%
  \draw[->] (regp2.west) --node[above]{$b_i$} (Pp.east);
  \draw[->] (meterp.south) --node[right]{$m'_i$} (Pp.north);
  
  \draw (-6,-1) node[draw,minimum width=1cm,fill=white](regv2){\scriptsize{Register $b$}};
  \draw (-3.5,-1) node[draw,minimum width=1cm,fill=white](Dv) {\scriptsize{Decoder}};%
  \draw[->] (regv2.east) --node[above]{$b_i$} (Dv.west);
  \draw[->] (Pp.west) --node[above,sloped]{{\tiny{\circled{4}}} $\Ket{m'_i}_{b_i}$} (Dv.east);
  \draw ($(Dv.east)+(0.5, .55)$) node {\begin{tabular}{l}\scriptsize{stop}\\\scriptsize{CLK}\end{tabular}};%
  
  \node[decision, below = .5cm of meterv] (compv) {$\stackrel{?}{=}$}; 
  \draw[->] (meterv.south) --node[right]{$m_i$} (compv.north);
  \draw[->] (Dv.north) --node[right]{$m'_i$} (compv.south);
  
  \draw[dotted] (-7,-1.75) -- (-2,-1.75);
  \draw[dotted] (1.25,-1.75) -- (5.75,-1.75);
  
    %Redefine roles: A=P , B=V
  \node[squarednode] at (-6,-2.5) (Nv) {\scriptsize{$A \leftarrow$ Prover}};
  \node[squarednode] at (5,-2.5) (Np) {\scriptsize{$B \leftarrow$ Verifier}};

  %Verifier to Prover DB
  
  \node at (3,-3.1) {For $j=1,2,\cdots,n$};
  
  \node[draw, ellipse, minimum height=1cm, minimum width=2cm] at (2.35,-4) (sourcep) {\scriptsize{Source}};
  \node[meter] (meterp1) at (2.35,-6) {};
  \draw[->] (sourcep.south) --node[right]{EP$_j$} (meterp1.north);
  
  \draw (4.85,-6) node[draw,minimum width=1cm,fill=white](regp3){\scriptsize{Register $a$}};
  \draw[->] (regp3.west) --node[above]{$a_j$} (meterp1.east);
  
  \node[meter] (meterv1) at (-3.5,-4.5) {};
  \draw[->] (sourcep.west) --node[above,sloped]{{\tiny{\circled{5}}} EP$_j$} (meterv1.east);
\draw ($(sourcep.west)+(0.2, -.85)$) node {\begin{tabular}{l}\scriptsize{start}\\\scriptsize{CLK}\end{tabular}};%
  \draw (-6,-4.5) node[draw,minimum width=1cm,fill=white](regv3){\scriptsize{Register $a$}};
  \draw[->] (regv3.east) --node[above]{$a_j$} (meterv1.west);
  
  \draw (-6,-8.5) node[draw,minimum width=1cm,fill=white](regv4){\scriptsize{Register $b$}};
  
  \draw (-3.5,-8.5) node[draw,minimum width=1cm,fill=white](Pv) {\scriptsize{Encoder}};%
  \draw[->] (regv4.east) --node[above]{$b_j$} (Pv.west);
  \draw[->] (meterv1.south) --node[right]{$m'_j$} (Pv.north);
  
  \draw (2.35,-9)node[draw,minimum width=1cm,fill=white](Dp) {\scriptsize{Decoder}};%
  \draw[->] (Pv.east) --node[above,sloped]{{\tiny{\circled{6}}} $\Ket{m'_j}_{b_j}$} (Dp.west);
  \draw ($(Dp.west)+(-0.2, .75)$) node {\begin{tabular}{l}\scriptsize{stop}\\\scriptsize{CLK}\end{tabular}};%
  
  \draw (4.85,-9) node[draw,minimum width=1cm,fill=white](regp4){\scriptsize{Register $b$}};
  \draw[->] (regp4.west) --node[above]{$b_j$} (Dp.east);
  
  \node[decision, below = .5cm of meterp1] (compp) {$\stackrel{?}{=}$}; 
  \draw[->] (meterp1.south) --node[right]{$m_j$} (compp.north);
  \draw[->] (Dp.north) --node[right]{$m'_j$} (compp.south);

  % Prover and Verifier dahsed boxes
  \draw (-4.5,1) node[draw,dotted,minimum width=5cm,minimum
  height=19cm]{} +(2,10)node{\begin{tabular}{c}Party A\\ $k$\end{tabular}}; %

  \draw (3.6,1) node[draw,dotted,minimum width=4.75cm,minimum
  height=19cm]{} +(1.8,10)node{\begin{tabular}{c}Party B\\ $k$\end{tabular}};

\end{tikzpicture}
  \caption{Entanglement-based one-way QDB protocol with reversed roles, to establish mutual distance bounds.% \textcolor{blue}{TODO: I think that denoting each step in this figures with a step number, and referencing those step numbers in the descriptions in the corresponding subsections (4.2.1)  will make the protocols clearer to understand.}\aysajan{I agree. Will do so.}
  }
  \label{Protocol1}
\end{center}
\end{figure*}
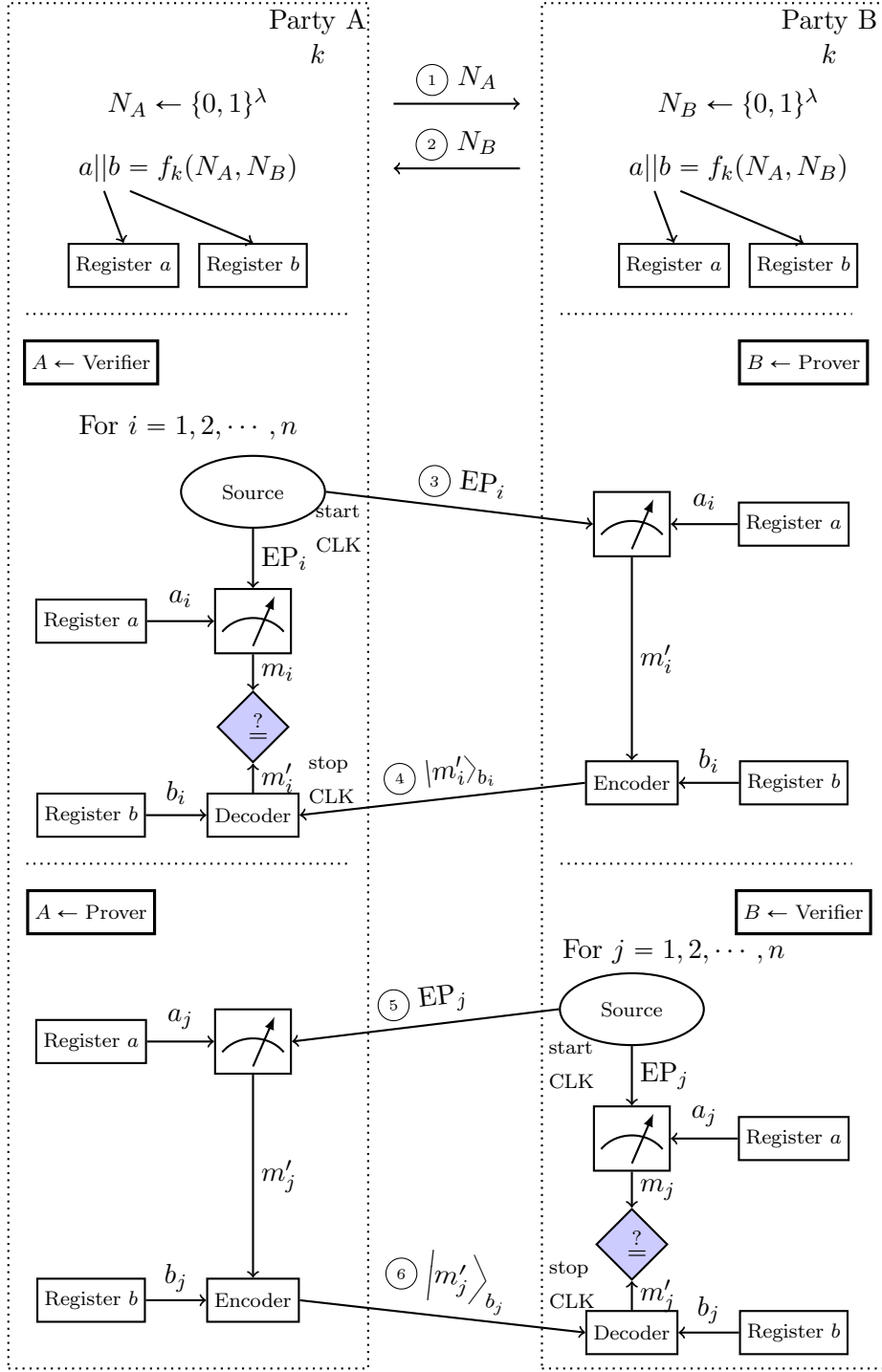

\subsubsection{Entanglement-based Mutual QDB \label{subsec:mutual_QDB_protocol}}

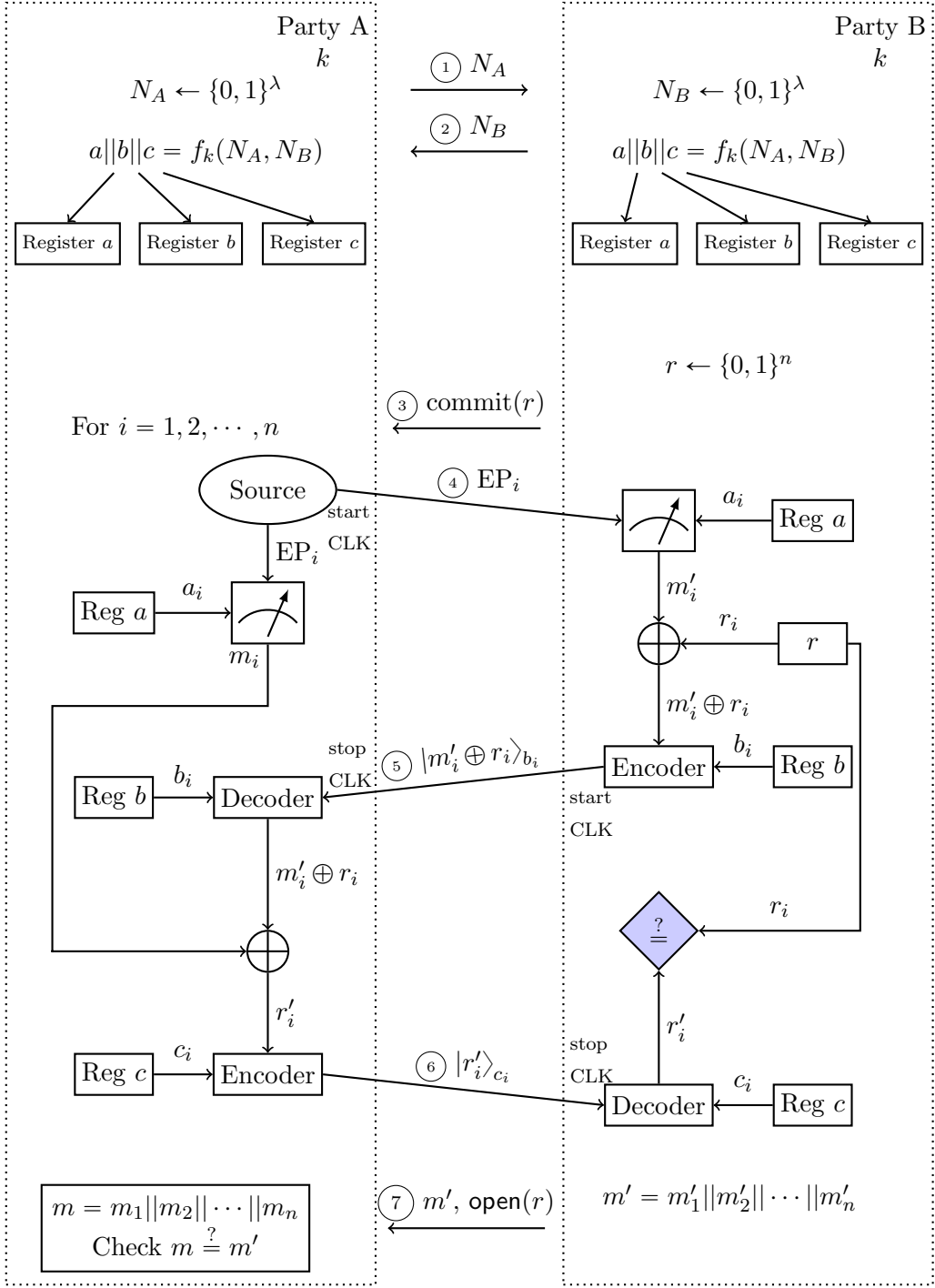
\begin{figure*}[hpt]
\begin{center}
\begin{tikzpicture}[scale=.9,thick,every node/.style={minimum height=.6cm}]
  \scriptsize
  %Initialisation phase
  \node[] at (-4.5,9.5) (Nv) {$N_A \leftarrow\{0,1\}^\lambda$};
  \node[] at (4,9.5) (Np) {$N_B \leftarrow\{0,1\}^\lambda$};
  \draw[->] ($(Nv.east)+(1.95,0)$) -- node[above]{{\tiny{\circled{1}}} $N_A$}($(Np.west)+(-1.9,0)$);
  \draw[<-] ($(Nv.east)+(1.95,-1)$) -- node[above]{{\tiny{\circled{2}}} $N_B$} ($(Np.west)+(-1.9,-1)$);
  
  \node[] at (-4.5,8.5) (prf1) {$a||b||c = f_k(N_A, N_B)$};
  \node[] at (4,8.5) (prf2) {$a||b||c = f_k(N_A, N_B)$};
  
  \node[draw] at (2.3,7) (regpa) {\scriptsize{Register $a$}};
  \node[draw] at (4.3,7) (regpb) {\scriptsize{Register $b$}};
  \node[draw] at (6.3,7) (regpc) {\scriptsize{Register $c$}};
  
  \node[draw] at (-6.75,7) (regva) {\scriptsize{Register $a$}};
  \node[draw] at (-4.75,7) (regvb) {\scriptsize{Register $b$}};
  \node[draw] at (-2.75,7) (regvc) {\scriptsize{Register $c$}};
  
  \draw[->] ($(prf1.south) + (-1.5,0)$) to (regva.north);
  \draw[->] ($(prf1.south) + (-1.1,0)$) to (regvb.north);
  \draw[->] ($(prf1.south) + (-.7,0)$) to (regvc.north);
  
  \draw[->] ($(prf2.south) + (-1.5,0)$) to (regpa.north);
  \draw[->] ($(prf2.south) + (-1.1,0)$) to (regpb.north);
  \draw[->] ($(prf2.south) + (-.7,0)$) to (regpc.north);
  %Distance bounding phase
  
  \node[] at (4,5) (Rpp) {$r\leftarrow\{0,1\}^n$};
  \node[] at (-4.5,5) (Rppp) {};
  
  \draw[<-] ($(Rppp.east)+(2.9,-1)$) -- node[above]{{\tiny{\circled{3}}} commit($r$)} ($(Rpp.west)+(-1.9,-1)$);

  %Prover to Verifier DB
  
  \node[draw, ellipse, minimum height=1cm, minimum width=2cm] at (-3.5,3) (sourcev) {Source};
  \node[meter] (meterv) at (-3.5,1) {};
  \draw[->] (sourcev.south) to node[right]{EP$_i$} (meterv.north);
  
  \draw (-6,1) node[draw,minimum width=1cm,fill=white](regv1){Reg $a$};
  \draw[->] (regv1.east) --node[above]{$a_i$} (meterv.west);

  \node[meter] (meterp) at (2.85,2.5) {};
  \node at (-5,4) {For $i=1,2,\cdots,n$};
  \draw[->] (sourcev.east) -- node[above]{{\tiny{\circled{4}}} EP$_i$} (meterp.west);
  \draw ($(sourcev.east)+(0.2, -.6)$) node {\begin{tabular}{l}\scriptsize{start}\\\scriptsize{CLK}\end{tabular}};%
  
  \draw (5.35,2.5) node[draw,minimum width=1cm,fill=white](regp1){Reg $a$};
  \draw[->] (regp1.west) --node[above]{$a_i$} (meterp.east);
  
  \draw (5.35,0.5) node[draw,minimum width=1cm,fill=white](rp){$r$};
  
  \node[XOR] at (2.85,0.5) (xor1) {};
  \draw[->] (rp.west) --node[above]{$r_i$} (xor1.east);
  \draw[->] (meterp.south) to node[right]{$m'_i$}(xor1.north);
  
  \draw (5.35,-1.5) node[draw,minimum width=1cm,fill=white](regp2){Reg $b$};
  \draw (2.85,-1.5)node[draw,minimum width=1cm,fill=white](Pp) {Encoder};%
  \draw[->] (regp2.west) --node[above]{$b_i$} (Pp.east);
  \draw[->] (xor1.south) --node[right]{$m'_i\oplus r_i$} (Pp.north);
  \draw ($(Pp.west)+(-0.2, -.75)$) node {\begin{tabular}{l}\scriptsize{start}\\\scriptsize{CLK}\end{tabular}};%
  
  \draw (-6,-2) node[draw,minimum width=1cm,fill=white](regv2){Reg $b$};
  \draw (-3.5,-2) node[draw,minimum width=1cm,fill=white](Dv) {Decoder};%
  \draw[->] (regv2.east) --node[above]{$b_i$} (Dv.west);
  \draw[->] (Pp.west) -- node[above,sloped]{{\tiny{\circled{5}}} $\Ket{m'_i \oplus r_i}_{b_i}$} (Dv.east);
  \draw ($(Dv.east)+(0.45, .55)$) node {\begin{tabular}{l}\scriptsize{stop}\\\scriptsize{CLK}\end{tabular}};%
  
 % \draw (-6,-4.5) node[draw,minimum width=1cm,fill=white](rv){$r_V\leftarrow\{0,1\}$};
  
  \node[XOR] at (-3.5,-4.5) (xor2) {};
  
  \draw[->] (meterv.south) -- ($(meterv.south)+(0,-1)$) node[near start, left]{$m_i$} -| ($(meterv.south)+(0,-1)+(-3.5,0)$) |- ($(meterv.south)+(0,-1)+(-3.5,0)+(0,-4)$) |- (xor2.west);
  
  \draw[->] (Dv.south) --node[right]{$m'_i\oplus r_i$} (xor2.north);
  
  \draw (-3.5,-6.5) node[draw,minimum width=1cm,fill=white](Pv) {Encoder};%
  \draw[->] (xor2.south) --node[right]{$r_i'$} (Pv.north);
  
  \draw (-6,-6.5) node[draw,minimum width=1cm,fill=white](regv3){Reg $c$};
  
  \draw[->] (regv3.east) --node[above]{$c_i$} (Pv.west);
  
  \draw (2.85,-7)node[draw,minimum width=1cm,fill=white](Dp) {Decoder};%
  \draw[->] (Pv.east) -- node[above]{{\tiny{\circled{6}}} $\Ket{r'_i}_{c_i}$} (Dp.west);
  \draw ($(Dp.west)+(-0.2, .75)$) node {\begin{tabular}{l}\scriptsize{stop}\\\scriptsize{CLK}\end{tabular}};%
  
  \draw (5.35,-7) node[draw,minimum width=1cm,fill=white](regp4){Reg $c$};
  \draw[->] (regp4.west) --node[above]{$c_i$} (Dp.east);
  
  \node[decision, text width=1em, below = 1.5cm of Pp] (compp) {$\stackrel{?}{=}$};
  
  \draw[->] (Dp.north) --node[right]{$r_i'$} (compp.south);
  
  \draw[->] (rp.east) -- ($(rp.east)+(.2,0)$) -| ($(rp.east)+(.2,0)+(0,-3.5)$) |-node[above, near end]{$r_i$} (compp.east);

  \node at (4, -8.5) {$m'=m_1'||m_2'||\cdots||m_n'$};

  \draw[->] (1, -9) to node[above]{{\scriptsize{\circled{7}}} $m',\,\textsf{open}(r)$}(-1.55,-9);
  
  \node[draw] at (-5, -9) (finalcheck1) {\begin{tabular}{c}$m=m_1||m_2||\cdots||m_n$\\Check $m\stackrel{?}{=}m'$ \end{tabular}};

  % Prover and Verifier dahsed boxes
  \draw (-4.75,.5) node[draw,dotted,minimum width=5.4cm,minimum
  height=18.75cm]{} +(2.15,9.75)node{\begin{tabular}{c}Party A\\ $k$\end{tabular}}; %

  \draw (4.3,.5) node[draw,dotted,minimum width=5.4cm,minimum
  height=18.75cm]{} +(2.15,9.75)node{\begin{tabular}{c}Party B\\ $k$\end{tabular}};

\end{tikzpicture}
  \caption{An overview of our entanglement-based mutual QDB protocol. %\textcolor{blue}{TODO: I think that denoting each step in this figures with a step number, and referencing those step numbers in the d escriptions in the corresponding subsections (4.2.2)  will make the protocols clearer to understand.}
  }
  \label{Protocol2}
\end{center}
\end{figure*}

An schematic illustration of the  steps of our new mutual QDB protocol is provided in Figure \ref{Protocol2}. The protocol enables two parties to mutually establish an upper bound on the distance between them with a single execution of this protocol.

Similar to the one-way setting, during the initialization phase of this protocol, the two parties exchange two random nonces $N_A$ and $N_B$ and apply the keyed PRF $f_k$ to the exchanged nonces (step 1 and 2 in the figure). The result of the PRF is a random binary string that is then split into three parts which we denote as registers $a$, $b$, and $c$.
The protocol then proceeds as follows (steps 4 to 6 are repeated $n$ times):
\begin{enumerate}
\setcounter{enumi}{2}

    \item Party B generates a random bit sequence $r\leftarrow\{0,1\}^n$ and sends a commitment to $r$  to Party A. 
    
    \item In the $i$-th round of rapid-bit exchange, Party A generates a pair of entangled particles EP$_i$, which is one of those mutually orthogonal Bell states $\beta_{00},\,\beta_{01},\,\beta_{10},\,\beta_{11}$ from Section \ref{sec:entanglement} and sends one half, which we again denote as EP$_i$, to Party B. Party B eventually receives EP$_i$ and measures it in a basis determined by register $a_i$. Denote the measurement result as $m'_i$.
    
    \item Party B computes $m' \oplus r_i$ and encodes the result in a basis determined by register $b_i$. Then $\Ket{m'_i \oplus r_i}_{b_i}$ is then sent to Party A.
    
    \item Party A eventually receives $\Ket{m'_i \oplus r_i}_{b_i}$ from Party B, stops the clock, decodes it in a basis determined by register $b_i$, then XORs the result with the result of measuring its half of EP$_i$ a basis determined by register $a_i$. The result of the above step (which should be $r_i$ in an honest execution) is then encoded a basis determined by register $c_i$, and transmitted back to Party B.
    Party B receives this response, and stops the clock and computes an upper bound on the distance to Party A based on the time of flight. Party B then decodes the response from Party A in a basis determine by register $c_i$ and compares the result to $r_i$. If they are equal, then Party B knows that Party A was not cheating. Otherwise, it aborts the protocol. 
    
    \item In the last phase, Party B sends $m'$ which is concatenation of all $m_i'$ measured in the basis determined by register $a$ in step 4 of each round above, and opens the commitment to $r$. Party A eventually receives and checks if it is equal to the measurement value of EP$_i$. If this check fails, it aborts the protocol. 
    
\end{enumerate}

%\aysajan{Fix for Fig 4: Use commitment to $r_p$ in the initialisation phase, and then open the commitment in the last phase. On verifier side check if opened commitment is equal to $r'_p$ obtained from the XOR of m and decoded bit.}

%\aysajan{Use $P_1$ and $P_2$, instead of verifier and prover, as Karim suggested.}

%\aysajan{Make quantum components explicit in the figures.}

%\dave{Fig 4 has been fixed. Only thing that still needs to happen, is make the quantum components explicit.}

%\dave{In Fig4, I used the notation EP to denote an entangled photon.}

\section{Security Analysis} \label{SecurityAnalysis}
We first discuss a new property that our QDB protocols rely on: detection of reflected entangled particles that are not measured by a cheating prover. Next, we provide an (informal) analysis of the security of our entanglement-based QDB protocols against several attacks: (1) distance fraud attacks, (2) mafia fraud attacks, (3) terrorist fraud attacks, and (4) implementation attacks.
A  formal security analysis is beyond the scope of this work, and is left as future work.

\subsection{Reflection attacks\label{subsec:reflectionanalysis}}
In some of the attacks that are described later in this section, a malicious prover could try to reflect an entangled particle without measuring it first, and use the reflected particle as a standard quantum particle that the protocol expects as a response. Assuming\footnote{Without loss of generality, as the reasoning applies to other cases too.} that the entanglement is with respect to the polarization of the particles, if the verifier wants to ensure that the prover did not reflect back the entangled particle, the verifier can perform a joint measurement of its local entangled particle and the response it receives in the complementary basis, e.g., diagonal in the case of vertical and/or horizontal polarization. If the prover reflected the challenge entangled particle, then the outcome of this joint measurement will indicate an unusually high correlation, which should not be the case if a standard quantum particle was instead sent back. 
We note that the rationale behind such detection is along the same lines of the rationale typically used in entanglement-based quantum key distribution (QKD) in protocols~\cite{Honjo:08,Takesue:10,Ribordy_2000} analogous to BB84~\cite{BB84}, but where one typically expects high correlations in the normal case when no eavesdropper is present (no attack), and low correlations when such an eavesdropper is present (attack). In our case, high correlations are expected when there is an attack, and low correlations when none is present.
We think that the observation that entanglement-based QDB provides such detection properties for advanced distance fraud attack strategies is of independent interest and may be utilized in other QDB settings, e.g., distributed, hierarchical, and/or group settings.

\subsection{Distance Fraud Attack\label{subsec:distance-fraud-attack-analysis}} In this type of attacks, a dishonest prover (controlled by the adversary) attempts to shorten the distance computed by the verifier. A strawman distance fraud attack strategy is to attempt to predict the challenge bit, and send the response before receiving the actual challenge. This attack, which also applies to the entangled QDB protocols presented in this paper, succeeds with a probability of $2^{-n}$ for $n$ challenges, which can be made negligible by increasing $n$; we thus argue that this attack is  rendered ineffective for a large $n$.

We note that previous QDB protocols \cite{Abidin2019} were susceptible to a more advanced strategy for a distance fraud attack. In that advanced strategy, assuming in the $i$-th round of rapid-bit exchange $a_i = b_i$, then an adversary could effectively  shorten the computed distance by reflecting the incoming photon without decoding and re-encoding it. This strategy results in a prover saving some processing time (probably only a few nano-seconds though). In this case, the (shorter) distance computed by a verifier depends on exact length of the saved processing time (of the prover). The root causes of this attack are (1) the prover knowing the bits $a$ and $b$ in advance, and hence being able to identify the rounds where the condition $a_i = b_i$ holds, and (2) the feasibility of reflecting photons without the need to decode and re-encode them. The probability of success of this attack is $2^{-\texttt{HD}(a,b)}$, where $\texttt{HD}$ is the Hamming distance between $a$ and $b$.\\

\textbf{Entanglement-based One-way QDB:} We argue that the advanced distance fraud attack strategy discussed above no longer works in our entanglement-based QDB protocols. Recall that in our one-way QDB protocol (Figure \ref{Protocol1}), the verifier transmits its challenge to the prover via an entangled particle (step 3 in the protocol), and expects a \emph{standard (un-entangled) quantum particle} as a response from the prover (step 4 in the protocol). If the prover would instead reflect the entangled particle of step 3 back in step 4, then that entangled particle would collapse to a random bit when the verifier tries to decode it in the basis determined by value of the register $b$. Because the output of this decoding operation will be a random bit, this strategy would not offer any advantage compared to the default strawman distance fraud attack described above (with a success probability of $2^{-n}$). Moreover, as discussed above, the verifier could even detect this reflection attack by measuring the correlation between the challenge and the response.\\

%\dave{I changed the security discussion on using entanglement-based QDB protocols above. I have put the original text in a comment, so you can change back of to the original text if you would prefer that version.}

%%We argue that against our entanglement-based QDB protocols, such an advanced distance fraud attack strategy provides not better advantage compared to the default strawman distance fraud attack; this is because of the no-cloning theorem \cite{no-cloning} in quantum mechanics. A cheating prover \emph{cannot} copy an entangled challenge particle it receives, so the best it can do is reflect it without processing it, otherwise, the challenge will be destroyed, i.e., collapse to a random state which will be measured at the verifier. So there are two cases: (i) reflect the entangled challenge particle without measuring it, or (ii) attempt to clone it and destroy it. 
%%In case (ii) the response in step $4$ in the one-way QDB case (Figure \ref{Protocol1}) has a probability of $1/2$ of having matched the correct response of the challenge bit (since it will be measured as a random one at the verifier).
%%In case (i), since no measurement is performed at the prover, it has to guess a random bit to encode as a response in step 4 in the one-way QDB case (Figure \ref{Protocol1}). So in summary, in both cases a cheating verifier has $1/2$ probability of passing the verification, and thus after $n$ challenge bits the probability is $2^{-n}$.

\textbf{Entanglement-based Mutual QDB:} A similar reasoning as in the one-way QDB case holds for Party B performing a distance fraud attack in the mutual QDB protocol (Figure \ref{Protocol2}). So let us now consider the case where Party A tries to perform a distance fraud attack. Recall that both the challenge Party A receives (step 5 in the protocol) and the response it transmits (step 6 in the protocol) are encoded using a \emph{standard quantum particle}. When it performs the advanced attack strategy where it reflects the incoming photon -- without decoding and re-encoding it -- in round $i$, then Party A is successful in the following cases:
\begin{itemize}
	\item If $b_i \neq c_i$, then the output of the decoder in basis $c$ at Party B (step 7 in the protocol) will be random, since the photon will be decoded in the wrong basis. So the output will be correct with probability $1/2$.
	\item If $b_i = c_i$, then the response sent to Party B will only be correct if $m' \oplus r_B = r_B$. This condition obviously only holds when $m'=0$. However, recall from Sect. \ref{sec:entanglement} that the there is no information encoded in the state of entangled qubits. The information comes into existence only after the entangled qubits are measured. What this means is that the value $m'$ remains unknown to any party until later when measured. Therefore, Party A cannot enforce the condition that $m'=0$; it will hold with probability $1/2$. However, Party A could cheat by not using entangled particles in step 4 of the protocol. Instead, it could generate an un-entangled particle with $m'=0$. Party B cannot check that it received an un-entangled particle and will accept it as a valid challenge. As a result of using an un-entangled particle, Party A will be able to reflect all incoming challenges (step 5 of the protocol) in each round where $b_i = c_i$. This increases the success probability to $2^{-\texttt{HD}(b,c)}$.
\end{itemize}
In summary, our mutual QDB protocol offers asymmetric resilience against distance fraud attacks. Party B has no advantage compared to the default strawman distance fraud attack, while Party A does have an advantage.

\subsection{Mafia Fraud Attack} 
%\karim{The analysis of this attack in \cite{Abidin2019} claims that the best the attacker can do is\\ $max((2^{-\texttt{HD}(a,b)}, (5/8)^n)$, where $\texttt{HD}$ is the Hamming distance between the two strings $a$ and $b$. Given my analysis of the distance fraud attack above, I claim that the $2^{-\texttt{HD}(a,b)}$ part now becomes $(1/2)^n$ so the success probability becomes $max((1/2)^n, (5/8)^n)$, so only $(5/8)^n$. PLEASE DOUBLE CHECK! If my reasoning is correct, then we can just reuse the analysis producing the $(5/8)^n$ success probability of this attack from \cite{Abidin2019}.}

In this form of attack, an adversary either (1) guesses the verifier's challenges in advance, sends these random challenges to the prover to obtain the prover's responses (prior to receiving any challenges from the verifier), and then uses the prover's responses as the responses for the verifier's actual challenges, or (2) intercepts the verifier's challenges and responds using randomly chosen bits (encoded in random basis).  
In the first case (similar to the analysis in \cite{Abidin2019}), there are four scenarios to consider: (1.1) the attacker guesses the basis correctly but guesses a wrong challenge bit, (1.2) the attacker guesses both the basis and the challenge bit correctly, (1.3) the attacker guesses the basis incorrectly but the challenge bit correctly, and (1.4) the attacker guesses both the basis and the challenge bit incorrectly. In case (1.1), the prover's overall response will always be incorrect and the attack fails. In case (1.2), the attacker always wins. In case (1.3), with probability $1/2$ the prover's measurement results in the correct challenge bit, so the attacker wins with probability $1/2$. In case (1.4), again the prover's measurement results in the correct challenge bit with probability $1/2$, so the attacker wins with probability $1/2$. Overall, the attacker wins with probability $1/2$ in the strategy where it obtains the prover's responses in advance, and after $n$-rounds, the success probability is $(1/2)^n$.  

In the second case,  where the attacker responds to the verifier's challenges using randomly chosen basis, there are again four scenarios to consider: (2.1) the attacker measures the verifier's challenge in the correct basis and uses the correct basis to respond, (2.2) the attacker measures the verifier's challenge in the correct basis but uses the wrong basis to respond, (2.3) the attacker measures the verifier's challenge in the wrong basis but uses the correct basis to respond, (2.4) the attacker measures the verifier's challenge in the wrong basis and also uses the wrong basis to respond. 

In case (2.1), the attacker always wins. In cases (2.2) to (2.4), the attacker wins with probability $1/2$ in each case. Overall, the attacker wins with probability $5/8$ using this attack strategy of intercept and resend, or $(5/8)^n$ after $n$ rapid-bit exchange rounds. Note that the use of entangled particles does not change the success probability of mafia fraud attacks compared to the QDB protocol in \cite{Abidin2019}. This is as expected, as it is used to improve the efficiency of the mutual QDB protocol and offer stronger protection against an adversarial prover.

We note that the attacker can also employ a similar strategy to the one employed by a dishonest prover in the distance fraud attack. That is, the attacker can  just reflect the verifier's challenges. However, we have shown in Section \ref{subsec:reflectionanalysis} that this strategy is not better than the one where the attacker responds with random bits, due to the use of entangled particles. Hence, there is nothing to gain from employing it, as the success probability of this strategy is lower than $(5/8)^n$.

It is also interesting to note that the attack success probability and strategy is similar for the one-way as for the mutual QDB protocol. In both cases, the adversary chooses the party it wants to authenticate to in advance. This party will be the verifier in the attack strategy, and the other party will be the prover. The adversary then performs one of the attack strategies discussed above to falsely convince the verifier that it is the legitimate prover. When the verifier needs to authenticate to the prover, the adversary can ignore the verifier's responses.

\subsection{Terrorist Fraud Attack} 
Protocols resisting terrorist fraud attacks follow the blueprint in Figure \ref{fig:basic-db} but with the bits of $b$ derived using an encryption scheme that uses the bits $a$ as key to compute $b=Enc_a(k_p)$, i.e., their design is such that that revealing $a$ and $b$ reveals the long-term secret $k_p$. In its present form, our QDB protocols are not-resistant to terrorist fraud attacks, as no information on the long-term shared secret key $k_p$ can be derived from $a$ and $b$ because we assume that $f_k(.,.)$ is a PRF. If we retain $a = f_{k_p}(N_{v},N_{p})$ and make $b=Enc_a(k_p)$ then our protocols in Figure \ref{Protocol1} and \ref{Protocol2}, similar to the previous QDB protocols in \cite{Abidin2019,AbidinQDB}, become resistant to terrorist fraud attacks. 

%\karim{Aysajan: Please read the discussion above, and double check with the terrorist fraud attack discussion in section 4 in your Wisec'19 paper. Did I get the reasoning correct?}

\begin{table}[t]
  \centering
\caption{Comparison of our entanglement-based one-way and mutual QDB protocols with other classical and quantum DB protocols. Note that $a$, $b$, and $c$ are the registers in the respective protocols. %Where $\textsf{HD}(B,D)$ means the Hamming distance between strings B and D. \karim{Please explain briefly what B and D are, or refer to the section discussing them. Should we also extend the table to indicate which ones are classical and which ones are quantum, and which ones can only do one-way DB?} 
}
\vspace{5pt}
\begin{tabular}{ | l | c | c | }
\hline
&&\\
 & \textbf{Distance Fraud} & \textbf{Mafia Fraud} \\
 &&\\
 \hline
 &&\\
%  Abidin {\it et al.} & $\Big(\frac{1}{2}\Big)^n$ & $\Big(\frac{3}{4}\Big)^n$ & $\Big(\frac{3}{4}\Big)^n$ \\
% &&&\\
% \hline
% &&&\\
 Brands-Chaum \cite{BC93} (Classical) & $\left(\frac{1}{2}\right)^{n}$ & $\left(\frac{1}{2}\right)^n$ \\
 &&\\
 \hline
  &&\\
 Hanke-Kuhn \cite{Hancke:2005:RDB:1128018.1128472} (Classical)& $\left(\frac{3}{4}\right)^{n}$ & $\left(\frac{3}{4}\right)^n$ \\
 &&\\
 \hline
  &&\\
%  Abidin {\it et al.} & $\Big(\frac{1}{2}\Big)^n$ & $\Big(\frac{3}{4}\Big)^n$ & $\Big(\frac{3}{4}\Big)^n$ \\
% &&&\\
% \hline
% &&&\\
 QDB \cite{AbidinQDB} (Quantum one-way)& $\left(\frac{3}{4}\right)^{n}$ & $\left(\frac{3}{4}\right)^{n}$ \\
 &&\\
 \hline
  &&\\
%  Abidin {\it et al.} & $\Big(\frac{1}{2}\Big)^n$ & $\Big(\frac{3}{4}\Big)^n$ & $\Big(\frac{3}{4}\Big)^n$ \\
% &&&\\
% \hline
% &&&\\
 QDB \cite{Abidin2019} (Quantum one-way) & $\left(\frac{1}{2}\right)^{\textsf{HD}(a,b)}$ & $\max\Big(\left(\frac{1}{2}\right)^{\textsf{HD}(a,b)},\left(\frac{5}{8}\right)^n\Big)$ \\
 &&\\
 \hline
  &&\\
 Hybrid DB \cite{Abidin2020} (Hybrid one-way) & $\left(\frac{1}{2}\right)^n$ & $\left(\frac{3}{4}\right)^n$\\
 &&\\
 \hline
  &&\\
%  Abidin {\it et al.} & $\Big(\frac{1}{2}\Big)^n$ & $\Big(\frac{3}{4}\Big)^n$ & $\Big(\frac{3}{4}\Big)^n$ \\
% &&&\\
% \hline
% &&&\\
 \textbf{Our One-way QDB Protocol} & $\left(\frac{1}{2}\right)^{\textsf{HD}(a,b)}$ & $\left(\frac{5}{8}\right)^n$\\
 &&\\
 \hline
 &&\\
 \textbf{Our Mutual QDB Protocol} & $\left(\frac{1}{2}\right)^{\textsf{HD}(b,c)}$ & $\left(\frac{5}{8}\right)^n$\\
 &&\\
 \hline
\end{tabular}
 \label{tab:comparison}
\end{table}

\subsection{Implementation Attacks}
As noted in previous work \cite{Abidin2019}, one implementation attack that always should be considered in QDB is the photon number splitting (PNS) attack. 
The PNS attack applies when a transmitted (quantum) pulse has more than one particle/photon.
In QDB, if the verifier's challenge qubit is composed of multiple photons, then, the adversary could attempt to split the extra photons and only transmit the remaining single photon to the prover as noted in \cite{Abidin2019}. When PNS is applied to the closest previous (one-way) QDB protocols~\cite{Abidin2019}, it was argued that the adversary could not do anything further beyond waiting for the response to arrive from the prover, and that the response will encode the verifier's challenge bit in a new basis, which is unknown to the adversary. In conclusion, the PNS attack is not more effective than the mafia fraud in the QDB protocols of ~\cite{Abidin2019}; note that other previous QDB protocols  as opposed to the original protocol in \cite{AbidinQDB} where the PNS attack is mitigated by including a final authentication phase, where the prover sends all the received challenge bits to the verifier. The same analysis of PNS above applies to our entanglement-based QDB protocols. 

%%We leave it as open problem whether the fact that the randomness of the challenge bits even to the challenger results in any additional security guarantees.
% I commented out the last sentence of this subsection, as I'm not sure that this security argument really holds. In a terrorist fraud attack, the prover is malicious, so it can choose not to use entangled particles.

\subsection{Comparison}
In Table~\ref{tab:comparison}, we compare our entanglement-based one-way and mutual QDB protocols with the state-of-the-art QDB protocols \cite{AbidinQDB,Abidin2019,Abidin2020}, as well as the two classical DB protocols \cite{BC93,Hancke:2005:RDB:1128018.1128472} from which all other DB protocols are derived. We refer the interested reader to \cite{avoine2019security}, and the references therein, for a recent survey on DB protocols. We can see from the table that both of our protocols compare favorably to the QDB protocol in \cite{Abidin2019}.  

%\section{Open Issues and Future Work} \label{Future}
%This paper paves the way for several future research directions: (1) formal security analysis of entanglement-based (one-shot) mutual QDB, (2) experimental realization of such entanglement-based mutual QDB, (3) combining entanglement-based (mutual) QDB with QKD, and (4) QDB for group settings.

%\noindent A more detailed discussion of the above research directions is pushed to the appendix due to space constraints.

\section{Open Issues and Future Work} \label{Future}
This paper paves the way for several future research directions: (1) formal security analysis of entanglement-based mutual QDB, (2) experimental realization of such mutual QDB, (3) combining entanglement-based (mutual) QDB with Quantum Key Distribution (QKD), and (4) QDB for group settings.

\noindent \emph{(1) Formal Security Analysis:} As mentioned in Section \ref{SecurityAnalysis}, our security analysis is brief and informal. The objective of our analysis is to provide intuition and  evidence that our protocols are secure, but we do acknowledge that more rigour is required to establish more confidence in our protocols. We note that
the current literature is missing a detailed formal treatment of DB protocols in the quantum setting, i.e., formalizing required security guarantees using a game-based definition \cite{BR06} when the adversary has quantum capabilities. We think that this is one of the most pressing future research directions to consider, to lay out a rigorous formal foundation for this topic.

\noindent \emph{(2) Experimental Realization of (Mutual) QDB:} We argue that our protocols are practically feasible and can be implemented, and experimentally verified using existing QKD technology. Our arguments about practical feasibility are based on those in~\cite{AbidinQDB} because similar experimental setups for transmitting and measuring (entangled) particles in QKD (and other close experiments) may be directly applicable to our protocols, i.e., to prepare, transmit, and then measure the entangled particles during the rapid-bit exchange phase. Finally, QKD and equipment required for similar experimental setups are commercially available, and have been  tested and verified several time as reported in ~\cite{gehring2015implementation} and references therein.

\noindent \emph{(3) Combined (Mutual) QDB with Key Establishment:} 
Our protocols assume that the two parties already share a secret key $k$.
It would be interesting to explore combining QDB with a key establishment functionality to obtain distance-bounded mutually shared keys, similarly as was proposed in~\cite {Singelee07SPEUCS} using classical distance bounding protocols. It may be possible to combine QDB and QKD protocols to obtain this functionality. We think that this research direction is theoretically and experimentally interesting.

\noindent \emph{(4) QDB for Group Settings:} QDB has so far only been considered  in the context of a single prover and a single verifier. To the best of our knowledge, there has been no prior work on QDB in group settings, where a set of (quantum-capable) provers interact with a set of (quantum-capable) verifiers. In the classic setting, the need for group distance bounding (GDB)~\cite{gdb} is motivated by several practical scenarios such as group device pairing, location-based access control and secure distributed localization. It is currently unclear (at least to us) what a  Group QDB (GQDB) protocol would look like.

\section{Conclusion}\label{Conclusion}
In this paper we present two new Quantum Distance Bounding (QDB) protocols that utilize entangled particles to communicate random qubits in the rapid-bit exchange phase.  Because the blueprint of our base protocol for one-way DB utilizes entangled particles in the rapid-bit exchange phase, we were able to extend it to perform mutual QDB between two parties (prover and verifier) in a single execution. Previous QDB protocols had to be executed twice with those roles reversed in each execution. Our protocols eliminate the need for communication from the prover to the verifier in the last authentication phase (especially, in the one-way DB case), which was necessary in some previous QDB protocols. %\karim{double check}.
To the best of our knowledge, our entanglement-based QDB protocol is the first mutual QDB protocol proposed in the literature. 
Finally, we briefly discuss several future research directions that can benefit from our QDB protocols, e.g., generalizing QDB to group settings which is currently unaddressed in the literature. We note that in the classical case, (communication) efficient DB protocols for group settings build upon mutual DB protocols. We argue that our mutual QDB protocol now opens the door for constructing efficient QDB protocols for group settings.

%\section*{Acknowledgments} 
%Aysajan Abidin and Dave Singel\'ee are supported by XXX. 

\bibliographystyle{splncsnat}
\bibliography{db}

\end{document}